\documentstyle[aps,prb,multicol]{revtex}
\input{psfig}
\begin{document}
\title{Shot noise in diffusive conductors: A quantitative analysis of
electron-phonon interaction effects}
\author{Y. Naveh, D. V. Averin, and K. K. Likharev}
\address{Department of Physics and Astronomy, State University \\
of New York, Stony Brook, NY 11794-3800}
\date{\today }
\maketitle

\begin{abstract}
Using the 'drift-diffusion-Langevin' equation, we have quantitatively
analyzed the effects of electron energy relaxation via their interaction
with phonons, generally in presence of electron-electron interaction, on
shot noise in diffusive conductors. We have found that the noise power $%
S_I(\omega )$ (both at low and high observation frequencies $\omega $) drops
to half of its 'mesoscopic' value only at $\beta \gtrsim 100,$ where $\beta $
is the ratio of the sample length $L$ to the energy relaxation length $l_{%
{\rm ph}}$ (the latter may be much larger then the dephasing length). It
means in particular that at low temperatures the shot noise may be
substantial even when $L\sim 10^{-2}$ -- $10^{-1}$ cm, and the conductor is
'macroscopic' in any other respect.
\end{abstract}

\renewcommand{\Re}{{\rm Re}} \renewcommand{\[}{\begin{equation}} %
\renewcommand{\]}{\end{equation}} 
\newcommand{\bea}{\begin{eqnarray}}
\newcommand{\eea}{\end{eqnarray}}

\begin{multicols}{2}

Macroscopic metallic conductors, in which electrons are in local
equilibrium, do not exhibit non-equilibrium (``shot'') noise at low
voltages. On the other hand, short diffusive conductors (with length $L$
much shorter than the effective electron-phonon relaxation length $\,l_{{\rm %
ph}}=\sqrt{D\tau _{{\rm ph}}}$) do show this effect as soon as the average
energy $eV/2$ acquired from the electric field becomes larger than the
energy scale $T$ of equilibrium thermodynamic fluctuations \cite{Beenakker
92,Nagaev 92,Naveh 97}. Electron-electron scattering (which become
important at $L>l_{ee}=\sqrt{D\tau _{ee}}$) affect the shot noise intensity
only slightly at low frequencies\cite{Nagaev 95,Kozub 96}, though at
frequencies above the reciprocal Thouless time $\tau _T^{-1}=D/L^2$ their
effect is more substantial \cite{Naveh 98}. This low sensitivity of shot
noise to electron-electron scattering, as well as to screening \cite{Naveh
97}, may be readily understood: electron-electron interaction cannot drain
the energy supplied by the external electric field from the electron
subsystem, which is thereby ''overheated''. Shot noise may be considered as
a direct result of this deviation from equilibrium.

The effect of electron-phonon interaction is quite different: it may
drain extra energy from the electron subsystem, bring it closer to
local thermal equilibrium, and hence suppress the shot noise. The
analysis of these effects has been addressed in many recent works,
both analytically \cite{Beenakker 92,Nagaev 92,Nagaev 95,Shimizu
92+93,Landauer 93,Liu 94+94a,Buttiker 95} and using numerical
Monte-Carlo simulations\cite{Liu 97} (see also Ref.~\cite{Gurevich
95+96}). While describing the reduction of shot noise with growing
values of $ \beta \equiv L/l_{{\rm ph}}$, these works cannot provide
realistic values for the noise at $\beta \gtrsim 1.$ For example,
assuming a simplifying form of the electron-phonon interaction, it was
concluded in Ref.~\cite{Shimizu 92+93} that the low-frequency noise
behaves as $S_I\sim e^{-\beta ^2}$. On the other hand, a crude
partition of the sample into $L/l_{{\rm ph}}$ segments \cite{Beenakker
92,Liu 94+94a,Buttiker 95}, or, alternatively, the assumption of a
constant relaxation time\cite{Landauer 93}, lead to the conclusion
that the noise should vanish much slower, as $\beta ^{-1}$, since such
approaches neglect the dependence of the relaxation rate on the
excitation energy.  This dependence was taken into account by
Nagaev\cite{Nagaev 92,Nagaev 95}.  By asymptotically solving the
Boltzmann equation, he showed that for the case of zero temperature
and zero frequency, and in the limit $\beta \gg 1$, the noise
intensity behaves as $S_I\sim \beta ^{-2/5}$. However, the full
dependence of the noise on $\beta $, as well as the more realistic
case of finite temperatures and frequencies, has not yet been
studied. Such a study was the goal of this work. We have used the
'drift-diffusion-Langevin' equation \cite{Naveh 97,Naveh 98} to
calculate the current noise power as a function of the ratio $\beta $,
the temperature $T$, and $\omega $, for the two limiting values of the
ratio $\gamma =L/l_{ee}$.

The drift-diffusion-Langevin equation\cite{Naveh 97,Naveh 98} is based on
integration of the Boltzmann-Langevin equation\cite{Kogan 69} over the
electron energies. Its self-consistent solution with the Poisson equation
that accounts for screening in the system enables one to calculate noise
power at arbitrary frequencies\cite{Naveh 97}. The outcome of this approach
may be summarized as the following simple recipe: for a conductor with
uniform cross-section, the spectral density of current fluctuations at any
point $x$ (along conductor's length) is 
\[
\label{generalnoise}S_I(x;\omega ,T)=\frac{2A}{L^2}\int_{-\frac L2}^{\frac L2%
}|K(x,x^{\prime };\omega )|^2{\cal S}(x^{\prime };T)\,dx^{\prime }
\]
with $A$ the cross-sectional area. The local noise correlator ${\cal S}(x;T)$
is given by 
\[
\label{SS}{\cal S}(x;T)=2\sigma (x)\int_0^\infty dE\,f_s(E,x;T)\left[ 1-{f}%
_s(E,x;T)\right] 
\]
with $\sigma (x)$ the local conductivity and $f_s(E,x;T)$ the
momentum-symmetric part of the local (steady state) distribution function at
a total energy $E$. The response function $K(x,x^{\prime };\omega )$ equals
1 at zero frequency, but at finite frequencies it is dependent on the
specific geometry of the conductor and its electrodynamic environment,
always obeying the following sum rule: 
\[
\label{intK}\frac 1L\int_{-\frac L2}^{\frac L2}K(x,x^{\prime };\omega
)\,dx^{\prime }=1.
\]

In this work, we concentrate on the geometry which looks most
promising for experimental observation of the considered effects,
namely, a thin and long conductor close to a ground plane. For the
quantitative applicability of our results, thickness of the conductor
and its distance from the ground plane should be much smaller than
$L$, but its second transversal dimension (parallel to the ground
plane) may be arbitrary. (As a result the model is applicable, e.g., to
a two dimensional electron gas gated by a close electrode, which also
serves as the ground plane). In this geometry the response function
for noise current in the external electrodes is very simple\cite
{Naveh 98}:
\[
\label{Kegrpl}K^e(x^{\prime };\omega )=\kappa \frac L2\frac{\cosh (\kappa
x^{\prime })}{{\rm sinh}(\kappa L/2)}.
\]
Here $\kappa =\sqrt{-i\omega /D^{\prime }}$, $D^{\prime }=D+\sigma A/C_0$,
$C_0$ is the (dimensionless) linear capacitance between the conductor and
the ground plane, and $D$ is the diffusion coefficient. The
non-equilibrium distribution function $%
f_s(E,x)$ should be found by solving the stationary Boltzmann equation. At
this stage it is convenient to use dimensionless quantities $\xi =x/L$, $%
\varepsilon =E/eV$, and $t=T/eV$. In the diffusion limit ($l\ll L$), the
Boltzmann equation is 
\[
\label{Boltzmann}- (D / L^2) \frac{d^2f_s(\varepsilon ,\xi )}{d\xi ^2}%
=I(\varepsilon ,\xi )
\]
with $I(\varepsilon ,\xi )$ the collision integral.

We study here two limiting cases. In the first limit, electron-electron
scattering within the conductor is negligible, $\gamma =L/l_{ee}\ll 1$, so
the collision integral involves only scattering by phonons. For the
deformation potential scattering it may be presented as\cite{Gantmakher 87} 
\bea
\label{collint} \nonumber 
I(\varepsilon ,\xi ) & = & \frac 1{\tau _V}\int_0^\infty d\omega
\,\omega ^2\left\{ (1-f_s)\left[ f_s^{+}(1+N)+f_s^{-}N\right] \right. \\
& & \left. -f_s\left[
(1-f_s^{-})(1+N)+(1-f_s^{+})N\right] \right\} 
\eea
with $f_s\equiv f_s(\varepsilon ,\xi )$, $f_s^{\pm }\equiv f_s(\varepsilon
\pm \omega ,\xi )$, $N\equiv N(\omega )\equiv 1/[\exp (\omega /t)-1]$, and 
\[
\frac 1{\tau _V}=\frac 1{2\pi }\left( \frac{eV}{2\hbar k_Fv_s}\right) ^3%
\frac{mk_F^2\Xi ^2}{\hbar ^2\rho v_s},
\]
with $k_F$ the Fermi wavenumber, $m$ the effective mass, $\Xi $ the
deformation potential constant, $v_s$ the sound velocity, and $\rho $ the
mass density of the material. Since the voltage drops only across the
conductor, the distribution function must approach the equilibrium
distribution at the conductor-electrode interfaces. The boundary conditions
for Eq.~(\ref{Boltzmann}) are therefore 
\[
\label{bc}f_s(\varepsilon ,\mp 1/2)=f_0(\varepsilon \mp 1/2)\equiv \frac 1{%
1+\exp \left( \frac{\varepsilon \mp 1/2}t\right) }.
\]

In the second limit, electron-electron scattering is strong, $\gamma \gg 1$,
so the electrons are locally thermalized. The distribution function is then
given by\cite{Pothier 97}
\[
\label{fTe}f_s(\varepsilon ,\xi )=\frac 1{1+\exp \left[ \frac{\varepsilon
+\xi }{t_e(x)}\right] }.
\]
The equation for the electron temperature $t_e(x)=T_e(x)/eV$ in this limit
was obtained in Ref.\cite{Nagaev 95} by multiplying the Boltzmann equation
by $\varepsilon $ and integrating it with respect to energy. The
resulting equation reads 
\[
\label{diffeqTe}\frac{\pi ^2}6\frac{d^2\left[ t_e^2(\xi )\right] }{d\xi ^2}%
=24\zeta (5)\frac{L^2}{\tau _VD}\left( t_e^5(\xi )-t^5\right) -1,
\]
where $\zeta (5)\simeq 1.04$ is the Riemann zeta function.

$\tau _V$ which appears in equations~(\ref{collint},\ref{diffeqTe})
is the energy relaxation time of electrons with typical energy $eV$.
The relaxation length is then $l_{{\rm ph}}=\sqrt{D\tau _V}$. One can
immediately notice from equations (\ref{Boltzmann}--\ref{collint}) and (\ref
{bc}--\ref{diffeqTe}) that the dependence of $f_s(E,x)$ on the physical
variables of the problem $eV$, $T$, $L$, and $l_{{\rm ph}}$ in each of our
cases is only through the parameters $t=T/eV$ and $\beta =L/l_{{\rm ph}}$%
\cite{note tau_V}. From equations (\ref{generalnoise},\ref{SS},\ref{Kegrpl})
it is then seen that for a uniform conductor [$\sigma (x)=\sigma $] the only
additional parameter which affects the normalized noise value $\alpha
=S_I/2eI$ is $|\kappa |L=\sqrt{\omega \tau _T}$, with $\tau _T=L^2/D^{\prime
}$ the effective Thouless time. In particular, at fixed $\omega \tau _T$ the
only dependence of the noise on the sample length is due to its ratio $\beta 
$ with the thermalization length.

Equation~(\ref{Boltzmann}) can be solved analytically in three limiting
cases: $\beta =\gamma =0$ (no energy relaxation), $\beta =0,\gamma
\rightarrow \infty $ (strong local thermalization), and $\beta \rightarrow
\infty $ (strong thermalization to lattice temperature). The corresponding
distribution functions are 
\begin{mathletters}
\label{limits}
\begin{eqnarray}
f_s(\varepsilon ,\xi ) &=&(1/2+\xi )f_0(\varepsilon 
+1/2) \nonumber \\
& &  +(1/2-\xi
)f_0(\varepsilon -1/2)\hspace{0.4cm}\beta =\gamma =0,  \label{limitsa} \\
f_s(\varepsilon ,\xi ) &=&\left\{ 1+\exp \left[ \frac{\varepsilon +\xi }{%
t_h(\xi)}\right] \right\} ^{-1}\hspace{0.4cm}\beta =0,\,\,\gamma \rightarrow
\infty ,  \label{limitsb} \\
f_s(\varepsilon ,\xi ) &=&\left[ 1+\exp \left( \frac{\varepsilon +\xi }t%
\right) \right] ^{-1}\hspace{0.4cm}\beta \rightarrow \infty ,  \label{limitsc}
\end{eqnarray}
with the hot-electron temperature $t_h(\xi)=\sqrt{t^2+3(1-4\xi
^2)/4\pi ^2}$. The 
frequency and temperature dependences of the noise power in these limiting
cases were given in \cite{Naveh 98}.

In order to study the crossover region ({\it i.e.}, at finite values of $%
\beta $) we solve equations (\ref{Boltzmann}) and (\ref{diffeqTe})
numerically. Results for the distribution functions for several values of $%
\beta $ are presented in figures 1 and 2. Figure~1 shows the dependence of $%
f_s(\varepsilon ,\xi )$ on energy and position at $t=0.01$, with no
electron-electron scattering. Plots (a) and (f) correspond to the limiting
cases described by Eqs. (\ref{limitsa}) and (\ref{limitsc}), respectively.
At large $\beta $ the integral in Eq.~(\ref{collint}) must be small, and so
the distribution is close to a Fermi Dirac distribution in most of the
sample. However, the effective temperature of this distribution approaches
the lattice temperature $t=0.01$ only at $\beta \gtrsim 10^5$. This form of
the distribution function means that a perturbation expansion of Eq.~(\ref
{Boltzmann}) around the distribution at $\beta \rightarrow \infty $ gives
results which are far from reality for values of $\beta $ smaller than
$\sim 10^5$. 
Note also that at distances of the order $L/\beta ^{1/2}$ from the edges of
the sample the distribution is much sharper than in the bulk. We will show
that this fact has strong implications on the high frequency noise.
\begin{figure}[tb]
\vspace{1.0cm}
\centerline{\hspace{95pt} \psfig{figure=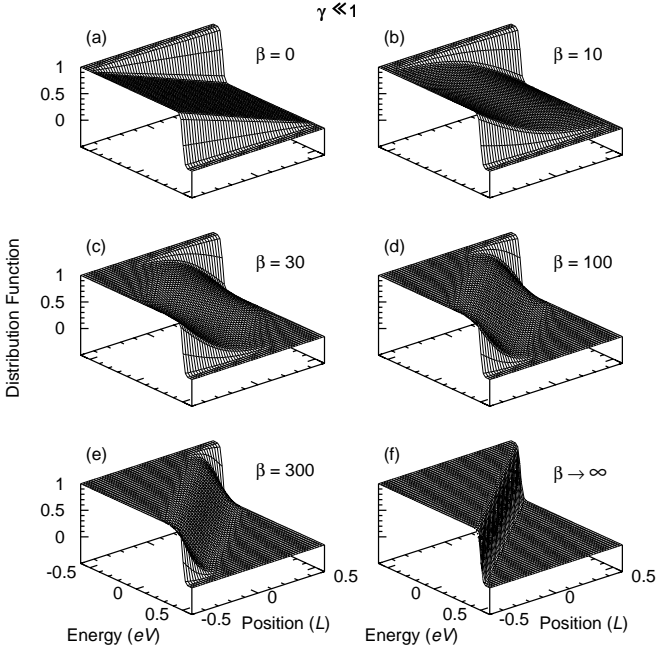,angle=-90,width=120mm}}
\narrowtext
\vspace{0.5cm}
\caption{The symmetric part of the electron distribution function
$f_s(\varepsilon 
,\xi )$ for different values of $\beta =L/l_{{\rm ph}}$ and for weak
electron-electron interaction $(\gamma =L/l_{ee}\ll 1)$. For all curves, $%
T=0.01eV$.}
\label{1noee}
\end{figure}

\begin{figure}[tb]
\centerline{\hspace{150pt} \psfig{figure=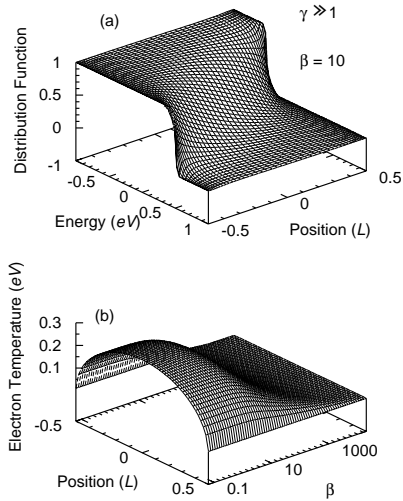,angle=-90,width=110mm}}
\narrowtext
\vspace{0.2cm}
\caption{(a) Same as in Fig.~1 but for the case $\gamma \gg 1$. (b)
The dependence 
of the electron temperature $t_e$ on the position and on the
electron-phonon 
interaction parameter $\beta $.}
\label{2strongee}
\end{figure}

Figure~2(a) shows the distribution function for the case of strong
electron-electron scattering, and $\beta =10$. At any other $\beta $, the
plot looks generally the same, with the only difference being that the width
of the distribution on the diagonal $\varepsilon +\xi =0$ would change
according to the local electron temperature $t_e(\xi ;\beta )$. The
dependence of this temperature on position and on $\beta $ is shown in
Fig~2(b). 

Figure 3 shows our main result: the normalized spectral density as a
function of $\beta $ for various frequencies. At $\omega \tau_T
,t\ll 1$ and $\beta \gg 1,$  Nagaev's asymptotic results\cite{Nagaev
92,Nagaev 95} are reproduced; in our notation they read 
\end{mathletters}
\begin{mathletters}
\label{asymptot}
\begin{eqnarray}
S_I/2eI &\simeq &\frac{1.2}{\beta ^{2/5}}{\,\,\,\,\,\,\,\,}(\gamma \ll 1),
\label{asymptota} \\
S_I/2eI &\simeq &\frac{1.05}{\beta ^{2/5}}{\,\,\,\,\,\,}(\gamma \gg 1).
\label{asymptotb}
\end{eqnarray}
However, for the case $\gamma \ll 1$ the result (\ref{asymptota})
relies on the fact that at 
large $\beta $ the electron distribution function is invariant to diagonal
transformations $\xi \rightarrow \xi +\xi _0,\varepsilon \rightarrow
\varepsilon -\xi _0$. As mentioned above [cf. Fig.~1(e)], this invariance is
not valid at distances $\xi \sim \beta ^{-1/2}$ from the edges of the
sample. Due to the form of the response function (\ref{Kegrpl}) the high
frequency noise is sensitive to the distribution of electrons only at
distances $\xi \sim 1/|\kappa L|=(\omega \tau _T)^{-1/2}$ from the edges.
Therefore, the noise behaves with its asymptotic power-law form only at $%
\beta \gg \omega \tau _T$, as clearly seen in Fig.~3 (solid lines).

\begin{figure}[tb]
\vspace{1.0cm}
\centerline{\hspace{20pt} \psfig{figure=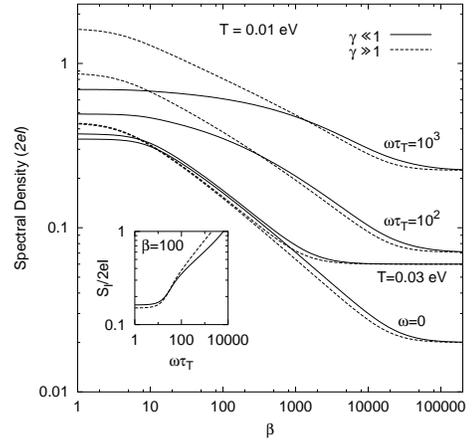,angle=-90,width=80mm}}
\narrowtext
\vspace{0.2cm}
\caption{The dependence of the noise spectral density on $\beta $ for
$t\equiv 
T/eV=0.01$ at various frequencies, for the two cases: $L\ll l_{ee}$ (solid
lines) and $L\gg l_{ee}$ (dashed lines). The same dependence for $t=0.03$
and zero frequency is also shown. Inset: the dependence of the spectral
density on frequency at $t=0.01$ and $\beta =100$.}
\label{3noise}
\end{figure}

A similar analysis can be done for the case $\gamma \gg 1$ (dashed lines in
Fig.~3). Here $S_I\sim \beta ^{-2/5}$ because $f_s(\varepsilon ,\xi )$ does
not depend on $\xi $ at $\beta \gg 1$. As can be deduced from Eq.~(\ref
{diffeqTe}), this constancy of $f_s$ does not hold at distances smaller than 
$\xi \sim \beta ^{-2/5}$ from the boundaries, and so, by the same arguments
as for $\gamma \ll 1$, the high frequency noise behaves as Eq.~(\ref
{asymptotb}) only at $\beta \gg (\omega \tau _T)^{5/4}$. In contrast
to the case of $\gamma \ll 1$, however, in this case the noise at
$1 \ll \beta < (\omega \tau _T)^{5/4}$ does not retain its $\beta =0$ value, 
because the electron temperature decreases with $\beta $ also near the
boundary, see Fig.~2(b). The inset of Fig.~3 shows the frequency-dependence
of the noise spectral density at $\beta =100$ and $t=0.01$. More details of
the dependence of the noise on frequency and temperature at $\beta =0$ were
given elsewhere\cite{Naveh 98}.

To summarize, we have performed numerical calculations of the
non-equilibrium noise in samples with strong elastic scattering and with an
arbitrary strength of electron-acoustic phonon scattering. We have shown
that the shot noise decreases very  slowly with the sample length, and
is of the order of 
the Schottky value even for $L/l_{{\rm ph}}\sim 100$. In experiment, it
means that when dealing with otherwise 'macroscopic' samples (at low
temperatures, of the order of one millimeter), the {\it a priori} assumption
of vanishing shot noise may be wrong. In view of the persistent improvements
in accuracy of noise measurements in recent years \cite{Liefrink 94,Reznikov
95,Steinbach 96,Henny 97,Schoelkopf 97} and the growing general interest in
the relations between inelastic relaxation, fluctuations, and dephasing \cite
{Mohanty 97}, we believe that there is a considerable chance of experimental
confirmation of this prediction.

The work was supported in part by DOE's Grant \#DE-FG02-95ER14575.

\end{mathletters}

\end{multicols}

\begin{references}
\bibitem{note tau_V}  This statement is correct only in the regular case
when the voltage is smaller than the Debye temperature $\Theta _D$.
Otherwise, the energy relaxation time does not depend on the voltage\cite
{Gantmakher 87}, and therefore $f_s$ depends separately on $V$ and $T$. The
case $\Theta _D\leq eV$ is not very practical due to the strong heating of
the lattice in this case.
\end{references}
\end{document}